\documentclass[a4paper,11pt,conference]{IEEEtran}

\usepackage{graphicx}  % Written by David Carlisle and Sebastian Rahtz

\usepackage{epstopdf}

\begin{document}

% paper title
\title{ Transferable tight binding model for strained group IV and III-V heterostructures }

\author{\authorblockN{%
Yaohua~P.~Tan, Michael~Povolotskyi, Tillmann~Kubis,
Timothy~B.~Boykin\authorrefmark{1} and Gerhard~Klimeck}
\authorblockA{Network for
Computational Nanotechnology, Purdue University, West Lafayette, Indiana, USA,47906}
\authorblockA{\authorrefmark{1}University of Alabama in Huntsville, Huntsville, Alabama, USA, 35899\\
e-mail: tyhua02@gmail.com}}

\maketitle

%%\begin{abstract}
%%\end{abstract}

%\section*{INTRODUCTION}
  Modern semiconductor devices have reached critical device
dimensions in the range of several nanometers. Devices such as
superlattice-FETs\cite{Pengyu_ED},Ultra Thin Body(UTB)-FETs and
FinFETs consist of  strained materials with different lattice
constant. Quantitative analysis of those devices requires the
reliable prediction of the bandgaps, effective masses in strained
heterostructures. The Empirical Tight Binding (ETB) methods are
appropriate for atomistic device modeling due to their numeric
efficiency\cite{Klimeck_ED}. However, the accuracy of ETB
calculations dependents on the transferability of the ETB
parameters. In this work, transferable ETB parameters of strained IV
and III-V group semiconductors are generated from \textit{ab-initio}
calculations\cite{Y_Tan_Mapping,Y_Tan_Mapping_prb}. The ETB
parameters show good transferability when applied to strained bulk
materials as well as ultra-thin superlattices.

%\section*{METHOD and RESULTS}
ETB parameters are obtained through \textit{ab-initio} mapping
process\cite{Y_Tan_Mapping_prb}. During the parameterization
process, ETB parameters and basis functions are adjusted to match
the corresponding \textit{ab-initio} band structures and wave
functions. In this work, \textit{ab-initio} bands of the strained
bulk materials and superlattices are calculated using VASP. Hybrid
functional HSE06 is used to produce correct band gaps. Group IV and
III-V materials are parameterized using the sp3d5s* ETB model. The
parameterized group IV and III-V materials include Si, Ge,
$\textrm{Si}_{0.5}\textrm{Ge}_{0.5}$ and compounds XY with X = Al,Ga
and In, Y = P, As and Sb. To have transferable ETB parameters,
following constraints are imposed. a) Onsites of each atom depend
only on the atom type instead of materials. b) Both strained and
unstrained ETB band structures are fitted to \textit{ab-initio}
results. c) Variation of interatomic coupling parameters among different materials is
less than 0.3eV. The strain effect is included using strain induced
onsite and interatomic couplings which depend on the change of bond
lengths of the first nearest neighbours and bond angles between the
first nearest neighbours.

\begin{figure}
\centering
\includegraphics[width=0.9\columnwidth]{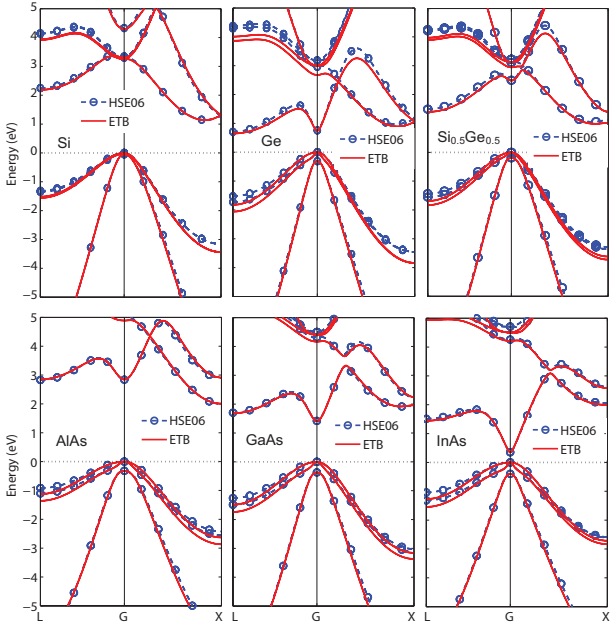}
\caption{Bulk band structure of Si, Ge, SiGe, AlAs, GaAs and InAs.
ETB bands shows good agreement with the hybrid functional
bands(HSE06).} \label{bulk_XAs_bands}
\end{figure}

\begin{figure}
\centering
\includegraphics[width=0.9\columnwidth]{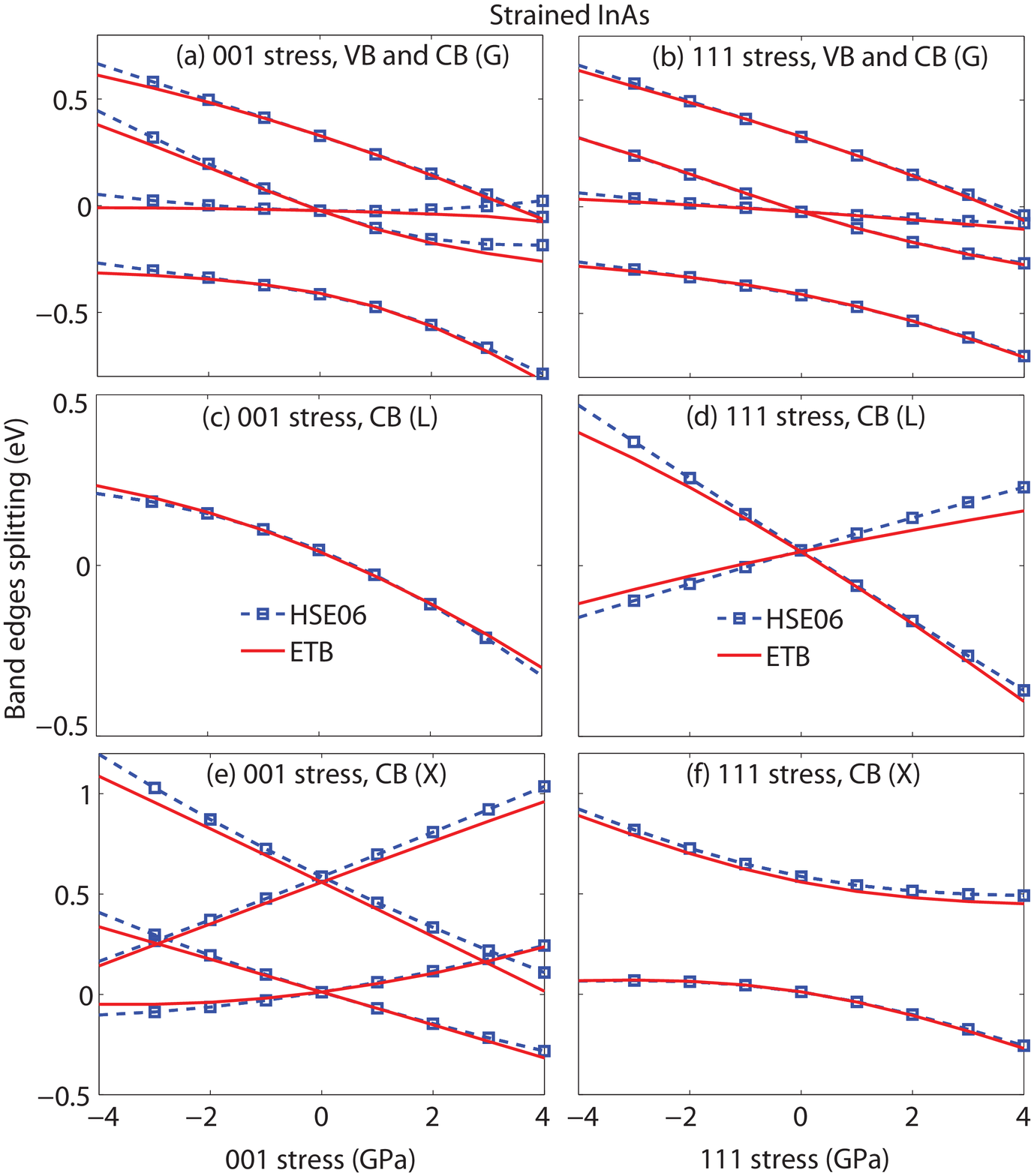}
\caption{Band edge at high symmetry points($\Gamma$,L,X) splitting
of strained InAs under 001 and 111 stress. The 0 energy of (a) and
(b) corresponds to the unstrained top valence bands; while the 0 energy of the
(c),(d) and (e),(f) correspond to lowest unstrained conduction band
at L and X points respectively. Compared with the HSE06 result, the
ETB model capture the band edge splitting at high symmetry point
under different strains.} \label{Strained_XAs_bands}
\end{figure}

\begin{figure}
\centering
\includegraphics[width=0.9\columnwidth]{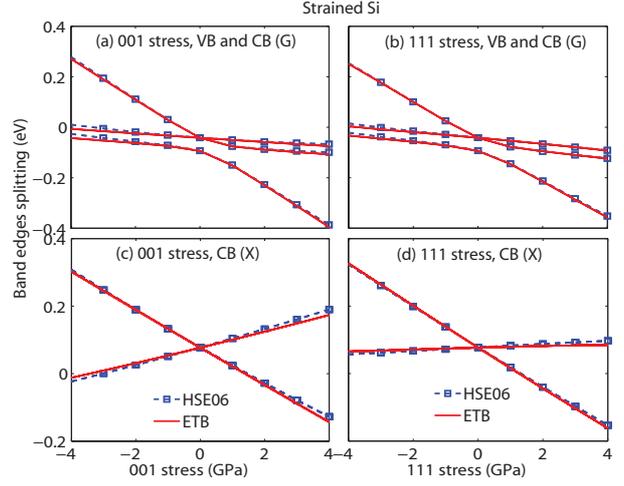}
\caption{Band edge at high symmetry points($\Gamma$,X) splitting
of strained Si under 001 and 111 stress. The 0 energy of (a) and
(b) corresponds to the unstrained top valence bands; while 0 of the
(c),(d) correspond to lowest unstrained conduction band
at X points.% Compared with the HSE06 result, the
%ETB model capture the band edge splitting at high symmetry point
%under different strains.
} \label{Strained_Si_bands}
\end{figure}

\begin{figure}
\centering
\includegraphics[width=0.9\columnwidth]{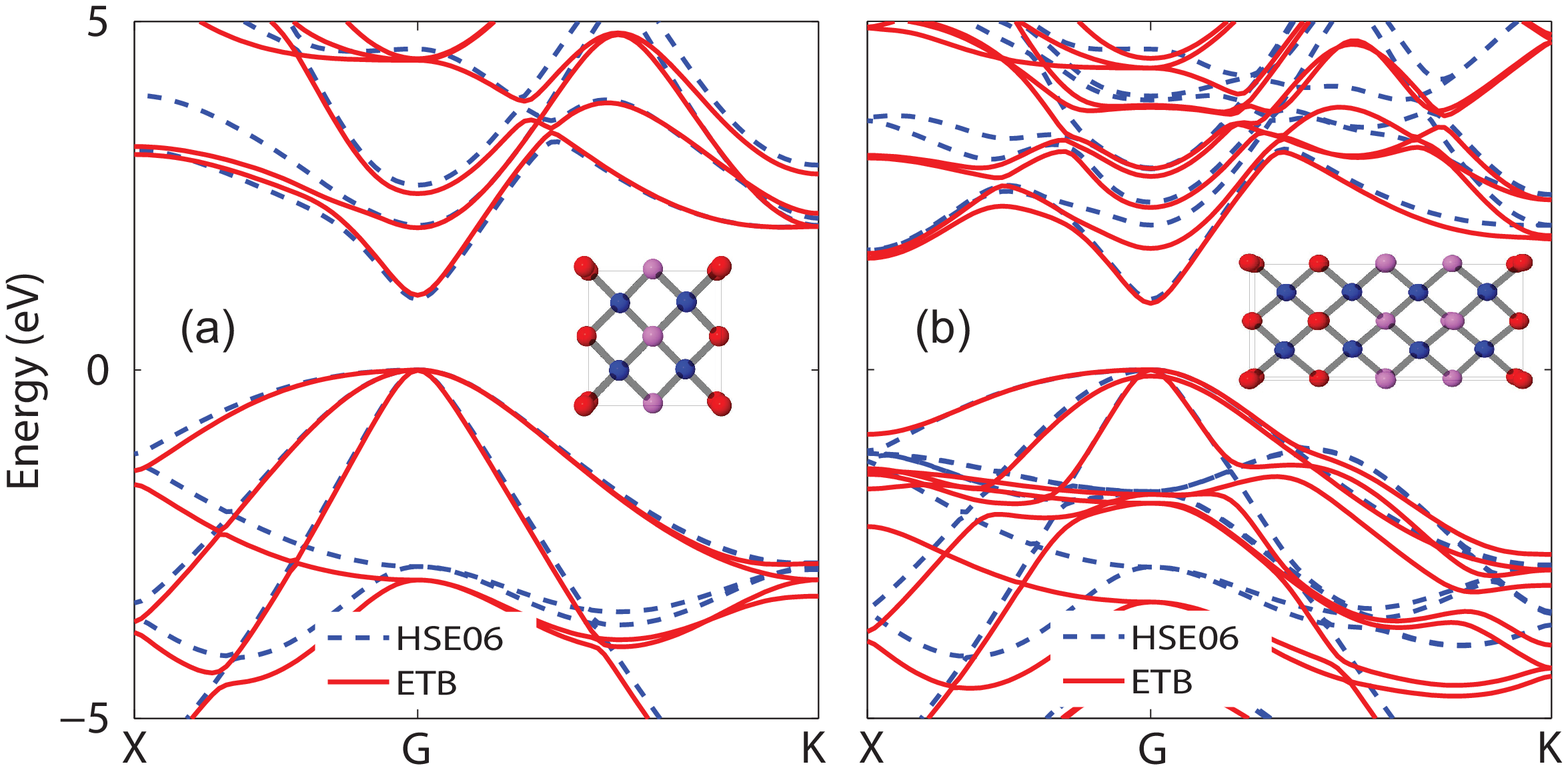}
\caption{Band structure of GaAs/InAs superlattices(HSE06 vs ETB).
Band structure of GaAs/InAs superlattices with 4 atomic layers (a)
and 8 atomic layers  (b) in the primitive unit cell. The ETB
parameters are transferable when applied to ultra thin
superlattices.} \label{Strained_XAsYAs_bands}
\end{figure}

\begin{figure}
\centering
\includegraphics[width=0.9\columnwidth]{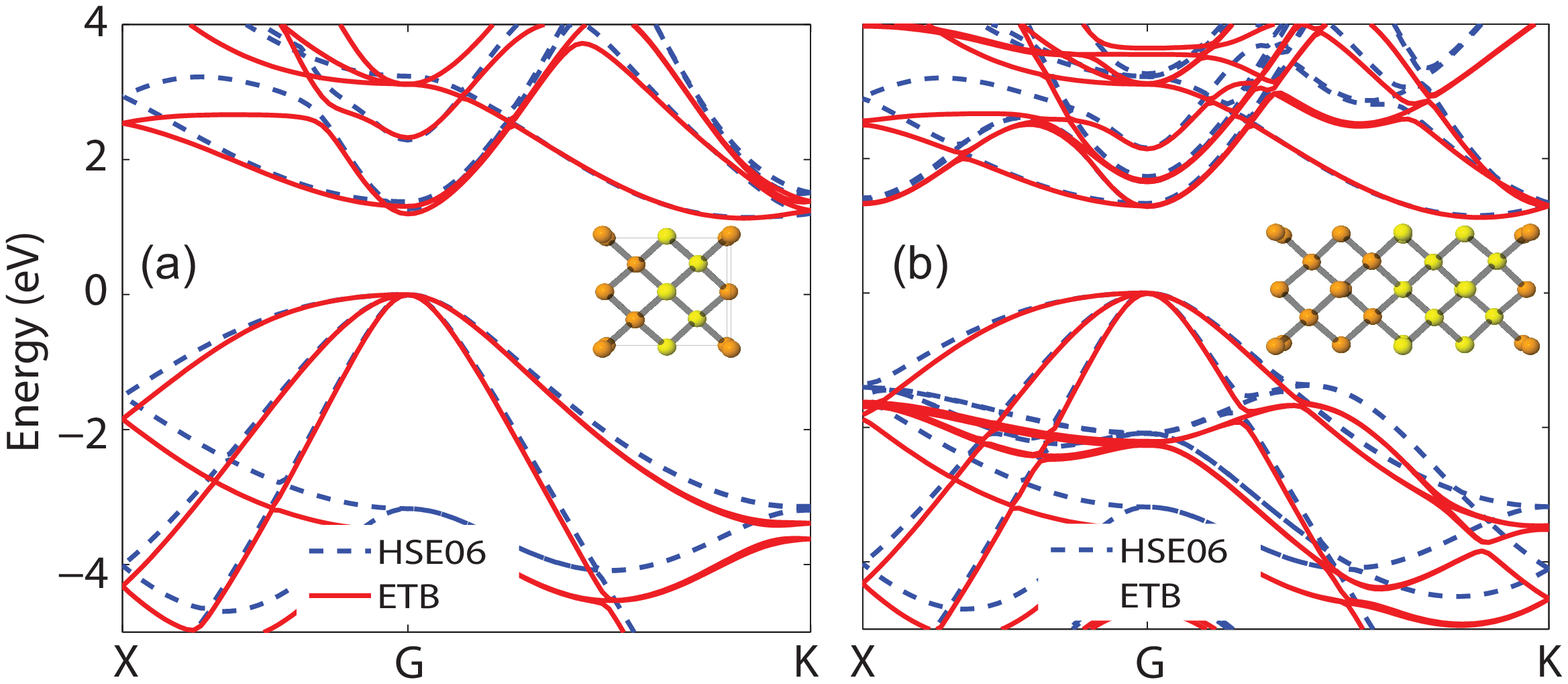}
\caption{Band structure of Ge/Si superlattices(HSE06 vs ETB). Band
structure of Ge/Si superlattices with 4 atomic layers (a) and 8
atomic layers  (b) in the primitive unit cell. The TB parameters are
transferable when applied to ultra thin superlattices.}
\label{Strained_GeSi_bands}
\end{figure}

Band structures of Selected semiconductors (bulk Si, Ge,
$\textrm{Si}_{0.5}\textrm{Ge}_{0.5}$, AlAs, GaAs and InAs) are shown
in Fig. \ref{bulk_XAs_bands}. The ETB band structures match the
corresponding hybrid functional calculations results well. Compared
with corresponding HSE06 results, band edge at high symmetry points
are within 0.05eV  and important effective masses have less than
10\% error. Fig. \ref{Strained_XAs_bands} and
\ref{Strained_Si_bands} show InAs and Si conduction and valence band
edge splitting under strains produced by stress along 001 and 111
directions. The splitting of conduction and valence band edges at
high symmetry points such as $\Gamma$, L and X are correctly
captured by the strain model. Fig. \ref{Strained_XAsYAs_bands} and
Fig. \ref{Strained_GeSi_bands} show the band structure of GaAs/AlAs
and Si/Ge superlattices respectively. The TB band structure agree
with the HSE06 bands, demonstrating good transferability of ETB
parameters for group IV and III-V semiconductors.

The use of nanoHUB.org computational resources operated by the Network
for Computational Nanotechnology funded by the US National Science Foundation
under Grant Nos. EEC-0228390, EEC-1227110, EEC-0228390, EEC-0634750, OCI-0438246, OCI-0832623 and OCI-0721680 is gratefully acknowledged.

%\begin{figure}
%\centering
%\includegraphics[width=0.9\columnwidth]{Fig5_Eg_vs_layer.eps}
%\caption{Band gap vs number of layers in supercells or GaAs/AlAs,
%GaAs/InAs and GaAs/AlAs superlattices.}
%\label{Strained_XAsYAs_bands}
%\end{figure}

% that's all folks
\end{document}